\documentclass[twocolumn,amsmath,amssymb,10pt,superscriptaddress,a4paper,letterpaper,fleqn]{revtex4}
\usepackage{amssymb}
\usepackage{epsfig}
\usepackage{graphicx}
\usepackage{dcolumn}
\usepackage{array}
\usepackage{bm}
\usepackage{fancyhdr}
\usepackage{longtable}
\usepackage{multirow}
\usepackage{float}
\usepackage{afterpage}
\usepackage{color}

\setlongtables
\usepackage[breaklinks=true,linkbordercolor={1 1 1}]{hyperref}

\begin{document}
\setcounter{page}{1}

\title{
\qquad \\ \qquad \\ \qquad \\  \qquad \\ 
Measuring light-ion production and fission cross sections versus elastic np-scattering at the upcoming NFS facility}
\author{K. Jansson}
\affiliation{Applied Nuclear Physics, Uppsala University, Sweden}
\author{C. Gustavsson}
\email[Corresponding author, electronic address:\\ ]{cecilia.gustavsson@physics.uu.se}
\affiliation{Applied Nuclear Physics, Uppsala University, Sweden}
\author{S. Pomp}
\affiliation{Applied Nuclear Physics, Uppsala University, Sweden}
\author{A.\,V. Prokofiev}
\affiliation{The Svedberg Laboratory, Uppsala University, Sweden}
\author{G. Scian}
\affiliation{Applied Nuclear Physics, Uppsala University, Sweden}
\author{D. Tarr\'io}
\affiliation{Applied Nuclear Physics, Uppsala University, Sweden}
\author{U. Tippawan}
\affiliation{Fast Neutron Research Facility, Chiang Mai University, Thailand}

\date{\today} 

\begin{abstract}
The Medley setup is planned to be moved to and used at the new neutron facility NFS where measurements of light-ion production and fission cross-sections are planned at 1-40\,MeV.  Medley has eight detector telescopes providing $\Delta E$-$\Delta E$-$E$ data, each consisting of two silicon detectors and a CsI(Tl) detector at the back. The telescope setup is rotatable and can be made to cover any angle. Medley has previously been used in many measurements at The Svedberg Laboratory (TSL) in Uppsala mainly with a quasi-mono-energetic neutron beam at 96 and 175\,MeV.

To be able to do measurements at NFS, which will have a white neutron beam, Medley needs to detect the reaction products with a high temporal resolution providing the ToF of the primary neutron. In this paper we discuss the design of the Medley upgrade along with simulations of the setup. We explore the use of Parallel Plate Avalanche Counters (PPACs) which work very well for detecting fission fragments but require more consideration for detecting deeply penetrating particles.
\end{abstract}
\maketitle

\lhead{Measuring light-ion production\ldots}
\chead{}
\rhead{K. Jansson \textit{et al.}}
\lfoot{}
\rfoot{}
\renewcommand{\footrulewidth}{0.4pt}

\section{INTRODUCTION}
Neutron For Science (NFS) is currently under construction. It will provide a white spectrum in the range 1-40\,MeV with competitive neutron flux \cite{NFS} from a full-stop Be target. The facility will offer flight paths of 5-25\,m where our plan is to use the shortest possible distance to benefit from the largest possible flux and statistics. 

We intend to measure both light-ion production as well as fission cross sections at NFS. The light-ion production in Si plays an important role in single-event upsets in electronics. Other elements are interesting because of their applications to dosimetry e.g. C and O present in biological matter or Fe, a common construction material. We also intend to use the Medley setup to measure fission cross sections as well as the energy-dependent anisotropy factor. $^{238}$U is important for neutron monitoring but also in nuclear power related areas, as well as for understanding the fission phenomena itself. At intermediate energies, above 20\,MeV, there are discrepancies among the reported cross-sections\,\cite{carlson}. Our intent is to make a precision measurement of the $^{238}$U(n,f) cross section but we plan to extend the measurements also to other actinides. The fission cross-section is planned to be measured relative the standard (n,p) cross section by simultaneous detection of protons at the forward angles.

\section{SETUP AND UPGRADE}
\begin{figure}[htb]
\includegraphics[width=0.4\columnwidth]{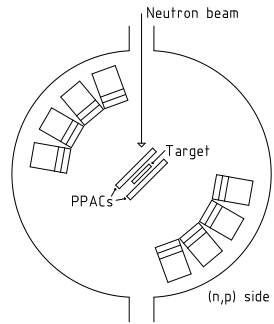}
\caption{The planned setup for the new fission measurements.}
\label{fig:Medley}
\end{figure}

The Medley setup which has been used earlier at 96 and 175\,MeV quasi-mono-energetic neutron energies \cite{Medley96,Medley175,Medley10y} will have to be upgraded to cope with the continuous energy spectrum. Medley consists of a chamber about 1\,m in diameter with eight detector telescopes at adjustable distances (typically $\sim$15\,cm from target) and viewing angles (Fig.\,\ref{fig:Medley}). In the existing setup each telescope provides $\Delta E$-$\Delta E$-$E$ data utilising one front and one middle Si-detector and one CsI(Tl)-scintillator in the back. Each telescope covers about 20\,msr. We will perform three main upgrades to facilitate our new measurements.

Since the neutron field at NFS will consist of neutrons of a continuous energy range we will have to determine the neutron energy for each event by measuring the neutron time-of-flight (ToF). We do so by measuring the time between the production of the neutrons and the first detected reaction product. We then need to find the ToF of the product particle between the target and the detection point. The Si-detectors measure the energy of the fission fragment (FF) but that is not enough to deduce the FF ToF. One way of obtaining the ToF is to install detectors with high temporal resolution and low stopping power close to the target. Currently we plan to install parallel plate avalanche counters (PPACs) which satisfy both these conditions. One drawback is that the thin mylar foils which constitute the parallel plates cannot handle large differences in pressure on the inside of the PPAC compared to the outside. A typical gas pressure in the PPACs for this kind of experiment is a few mbar. Therefore we plan to fill the whole chamber with the same gas and pressure while making sure that the FFs do not lose too much energy during their flight. The PPACs still introduce material in the beam which will cause np-scattering and thereby increased background.

When we detect protons and other light ions, the problem is somewhat less complicated since the particle species are deduced using the $\Delta E$-$E$ technique. The particle velocity can be calculated based on the total energy deposit in the telescope and the particle mass. In this way the telescopes can provide the particle ToF by themselves and therefore also the neutron ToF. If we can get sufficient temporal resolution from the Si-detectors we do not need any extra detectors for the light-ion detection. This would be beneficial since the PPACs probably will not be able to detect low-ionising particles with high efficiency.

As another upgrade, we intend to replace the front Si-detector with a thinner one, 25\,$\mu$m thick. This will allow us better resolution in the low energy region for identifying the light ions as well as clear discrimination between $\alpha$ particles and FFs. Any FF will be fully stopped in the front detector but an $\alpha$ particle with $\gtrsim$5\,MeV will penetrate it and deposit the rest of its energy in the middle detector (and possibly in the CsI). Thus, any energy deposit higher than 5\,MeV can be identified as a FF. According to energy loss calculations and simulations the FF will reach the detector with at least 10\,MeV.

One additional modification is needed. The target is tilted 45$^\circ$ with respect to the beam to allow all telescopes both in the 0-90$^\circ$ range as well as in the 180-270$^\circ$ range a good view of the respective facing side of the target. To simultaneously measure relative the H(n,p) cross section we will deploy a layered target with a uranium deposit on both sides of some polyethylene. That way all telescopes can detect FF at the same time as the front telescopes also detect proton recoils (see Fig.\,\ref{fig:Medley}). We are also considering using a dual layer target\label{sec:dual}. Then, we only measure FFs in the backward directions and only need one PPAC. To clearly distinguish (n,p) events originating from hydrogen from those originating from carbon ($Q$-value of -12.6\,MeV) the smallest angle telescope will need to be positioned as close to the beam as possible.

The light-ion production target will still be singled layered. A polyethylene as well as a pure carbon reference target has to be measured separately.

\section{SIMULATIONS}
To predict how our future setup will perform we are simulating, using the Geant toolkit \cite{Geant4}, the whole chamber with target and detectors in reasonable detail. Most important parameters as target materials, beam width, whether or not to use PPAC etc. can be controlled through macros. The neutron source is positioned outside the chamber five meters away and directed towards the target. Due to the low probability of neutron interactions an interface for biasing has been implemented based on \cite{bias} to reduce the amount of computer time required.

\begin{figure}[htb]
\includegraphics[width=0.8\columnwidth]{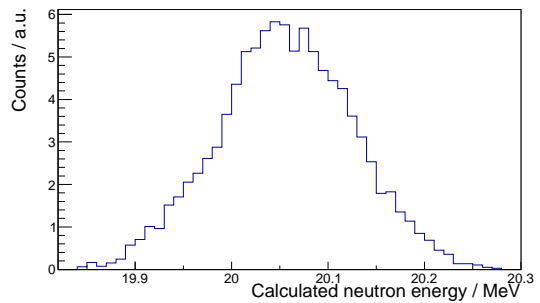}
\caption{Estimation of the incoming neutron energy using Eq.\,\ref{eq:tof} to determine the neutron ToF. An incoming mono-energetic beam of 20\,MeV neutrons was simulated.}
\label{fig:tof}
\end{figure}

Using mono-energetic incoming neutrons we can estimate how well we can determine the ToF using the timing signal in the PPAC and front Si-detector. To find the time $t_\text{ToF}$, when the neutron reached the target, the following formula, utilising the mean FF velocity, was used:
\begin{equation}
\label{eq:tof} t_\text{ToF} = t_\text{PPAC} - d_\text{PPAC}\times\frac{t_\text{Si}-t_\text{PPAC}}{d_\text{Si}-d_\text{PPAC}},
\end{equation}
where $t_x$ denotes the time of the detector signal from $x$ and $d_x$ similarly denotes the distance of $x$ from the target. $d_\text{PPAC}$ was arbitrarily set to 1\,cm. Calculating the neutron energy from the corrected ToF in Eq.\,\ref{eq:tof} gives a mean value of 20.06\,MeV (rms of 0.07\,MeV) for 20\,MeV incoming neutrons (Fig.\,\ref{fig:tof}). The higher calculated value is expected since the velocity between the target and the PPAC is higher than between the PPAC and the Si-detector due to energy losses. Without the second term in Eq.\,\ref{eq:tof} the mean value becomes 19.3\,MeV (rms of 0.2\,MeV).

\begin{figure}[htb]
\includegraphics[width=0.8\columnwidth]{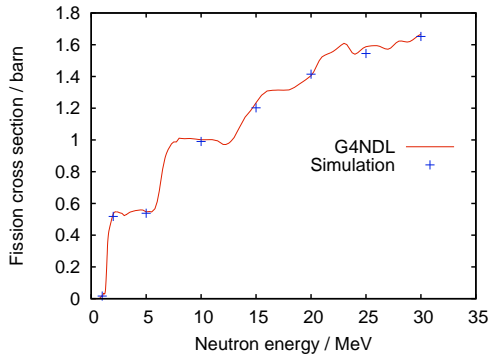}
\caption{Comparison of the $^{238}$U fission cross section deduced from the simulation with the same cross section taken from the nuclear data library G4NDL used in Geant4.}
\label{fig:G4NDL}
\end{figure}

Using the events collected in the backward telescopes (dual layer target, see Sec.\,\ref{sec:dual}) we have been able to reproduce the fission cross section (Fig.\,\ref{fig:G4NDL}) used in Geant4 (from G4NDL nuclear data library). This gives us some confidence that the simulations are giving a representative picture of our future experiment. A small correction, $\sim$1-3\%, due to the frame boost by the incoming neutron was needed and has been applied afterwards for each neutron energy.

\begin{figure}[htb]
\includegraphics[width=0.8\columnwidth]{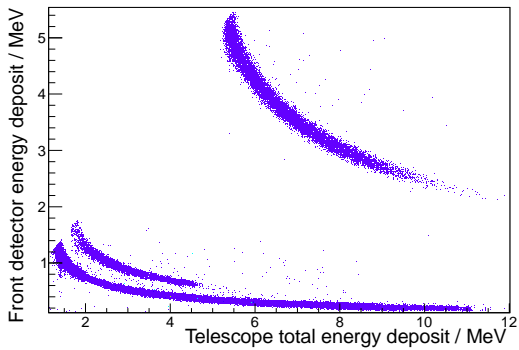}
\caption{Simulated $\Delta E$-$E$ plot showing light-ion production from a SiO$_2$ target. From top to bottom: the $\alpha$ particles, the deuterons and the protons can clearly be separated.}
\label{fig:dee}
\end{figure}

For the light-ion experiments the simulations are made without PPACs and therefore also with vacuum in the Medley chamber. Depicted in Fig.\,\ref{fig:dee} is one of the simulated $\Delta E$-$E$ plots that can be used for identifying the light-ions hitting a telescope.

\section{UNCERTAINTIES}
We intend to get a statistical uncertainty $<1\%$ which can be achieved for each telescope in $\sim $50\,h of beam time assuming a 1\,MeV bin size and that the average flux is about $10^6$/MeV/cm$^2$/s\,\cite{NFS}. The total incoming neutron flux cancels out for the fission measurement since we measure relative H(n,p) simultaneously with the same beam. Systematic uncertainties due to the differences in solid angle coverage are intended to be eliminated by rotating the telescopes 180$^\circ$. The uranium mass deposit will be determined by doing a long off-beam $\alpha$ decay measurement using the telescopes to count $\alpha$ particles. This should give us an accurate estimate of the number of uranium atoms. The H(n,p) reaction in the mylar foils of the PPAC is estimated to be one of the largest sources of uncertainty and we are currently working on the best way of minimising it. Another source of uncertainty is that the sensitive area of the Si-detectors will be different for protons compared to FF. This is believed to be a small effect but will be investigated in the future. Due to the closeness of the PPAC to the target, timing uncertainties will affect the FF measurement less drastically than the light-ion measurement.

For the light-ion experiment, but also for the (n,p) scattering part of the fission experiment, the Si-detectors' temporal resolution is probably the most critical. With 5\,m neutron flight path we need a temporal resolution in the order of a few ns to determine the neutron energy within 1\,MeV for 20\,MeV neutrons. The theoretical limit is estimated to be less than 1\,ns but our true resolution will be worse. The beam particle time spread on the neutron production target is reported to be less than one ns\,\cite{NFS}.

Our aim is a total uncertainty of 2\% for the fission measurement relative H(n,p). Regarding the light-ion measurement, it is completely dependent on the Si-detectors' timing properties.

\section{CONCLUSIONS AND OUTLOOK}
We have ongoing laboratory tests to try to determine the best possible time resolution of the Si-detectors using digital techniques. The outcome of these investigations will greatly affect how we choose to detect light-ions with the Medley setup. If the resolution is shown to be insufficient we will have to look into using additional fast and thin detectors to obtain the neutron ToF. Unfortunately the PPAC is unlikely to work for light-ions because of the low energy deposition, however the PPAC seems to be the best choice for detecting FF without making them lose too much energy. We are about to start laboratory tests also on existing PPACs (lent by our colleagues at IPNO) to see what we can do with these detectors before we design and construct our own.

With the above mentioned upgrades Medley will be prepared to measure both fission and light-ion production by the time the construction of NFS is finished.

\end{document}